%% file: ridge_PRL_postRef.tex
\newcommand*{\ATLASLATEXPATH}{}
\pdfoutput=1
\documentclass[cernpreprint,USenglish,texlive=2011]{\ATLASLATEXPATH atlasdoc}
\usepackage[biblatex=false]{\ATLASLATEXPATH atlaspackage}
\usepackage{\ATLASLATEXPATH atlascontribute}
\usepackage{\ATLASLATEXPATH atlasphysics}

\usepackage{ridge_PRL-defs}
\usepackage[square, comma, numbers,sort&compress]{natbib}
\AtlasTitle{Observation of long-range elliptic azimuthal anisotropies in
$\sqrt{s}=$13 and 2.76~TeV $pp$ collisions with the ATLAS detector}

\AtlasRefCode{HION-2015-09}
\PreprintIdNumber{CERN-PH-EP-2015-251}

\AtlasJournalRef{Phys. Rev. Lett. 116, 172301}
\AtlasDOI{10.1103/PhysRevLett.116.172301}

\AtlasAbstract{%
ATLAS has measured two-particle correlations as a function of
relative azimuthal-angle, $\Delta \phi$, and pseudorapidity, $\Delta \eta$,
in $\sqrt{s}$=13 and 2.76~TeV $pp$ collisions at the LHC using charged particles measured 
in the pseudorapidity interval $|\eta|$<2.5. The correlation functions evaluated 
in different intervals of measured charged-particle multiplicity show a 
multiplicity-dependent enhancement at $\Delta \phi \sim 0$ that extends over a 
wide range of $\Delta\eta$, which has been referred to as the ``ridge''. 
Per-trigger-particle yields, $Y(\Delta \phi)$, are measured over 2<$|\Delta\eta|$<5.  
For both collision energies, the $Y(\Delta \phi)$ distribution in all multiplicity 
intervals is found to be consistent with a linear combination of the per-trigger-particle 
yields measured in collisions with less than 20 reconstructed tracks, and a constant
combinatoric contribution modulated by $\cos{(2\Delta \phi)}$. The fitted
Fourier coefficient, $v_{2,2}$, exhibits factorization, suggesting that the ridge 
results from per-event $\cos{(2\phi)}$ modulation of the single-particle distribution 
with Fourier coefficients $v_2$. 
The $v_2$ values are presented as a function of multiplicity and transverse 
momentum. They are found to be approximately constant as a function 
of multiplicity and to have a $p_{\mathrm{T}}$ dependence similar to that measured 
in $p$+Pb and Pb+Pb collisions. The $v_2$ values in the 13 and
2.76~TeV data are consistent within uncertainties. 
These results suggest that the ridge in $pp$ collisions 
arises from the same or similar underlying physics as observed in $p$+Pb collisions, and 
that the dynamics responsible for the ridge has no strong $\sqrt{s}$ dependence.
}
\hypersetup{pdftitle={ATLAS draft},pdfauthor={The ATLAS Collaboration}}

\newcommand{\papertype}{letter}

\begin{document}

\maketitle

Measurements of two-particle angular correlations in high-multiplicity proton-proton (\pp) 
collisions at a center-of-mass energy $\sqs = 7$~\TeV\ at the LHC showed an enhancement in the production of pairs at small 
azimuthal-angle separation, \dphi, that extends over a wide range of pseudorapidity 
differences, \deta,  and which is often referred to as the ``ridge''~\cite{Khachatryan:2010gv}. The ridge has also been observed in proton-lead (\pPb) 
collisions~\cite{CMS:2012qk,Abelev:2012ola,Aad:2012gla,Chatrchyan:2013nka,Aad:2014lta,
Khachatryan:2015waa}, where it is found to result from a global sinusoidal modulation 
of the per-event single-particle azimuthal angle distributions~\cite{Abelev:2012ola,
Aad:2012gla,Chatrchyan:2013nka,Aad:2014lta}. While many theoretical interpretations of the ridge, including those 
based on hydrodynamics \cite{Bozek:2010pb,Werner:2010ss,Bozek:2012gr,Bzdak:2013zma,Bozek:2013uha}, saturation \cite{dumitru:2010iy,Ryskin:2011qh,
Dusling:2012iga,Tribedy:2011aa,Dusling:2012wy,Dusling:2013oia,Dumitru:2014vka,
Noronha:2014vva,Kovchegov:2013ewa,Dumitru:2015cfa,Lappi:2015vha}, or other mechanisms \cite{d'Enterria:2010hd,Avsar:2010rf,Levin:2011fb,
Kovner:2012jm,Diehl:2011tt,Bjorken:2013boa,Gyulassy:2014cfa}, have been, or could be 
applied to both \pp\ and \pPb\ collisions,
it has not yet been demonstrated that the ridge in \pp\ 
collisions results from single-particle azimuthal anisotropies.
Testing whether the ridges in \pp\ and \pPb\ 
collisions arise from the same underlying features of the single-particle distributions 
may provide insight into the physics responsible for the phenomena. Separately, a study
of the \sqs\ dependence of the ridge in \pp\ collisions may help distinguish between 
competing explanations. 

This \papertype\ uses 14~\inb\ of $\sqs = 13$~\TeV\ data and 4.0~\ipb\
of $\sqs=2.76$~\TeV\ data recorded during LHC Run~2 and Run~1,
respectively, to address these issues. The maximum number of inelastic interactions per
crossing was 0.04 and 0.5 for the 13 and 2.76~\TeV\ data,
respectively.  Two-particle angular correlations are measured as a function
of \deta\ and \dphi\ in different intervals of the measured
charged-particle multiplicity and different \pT\ intervals spanning
0.3<$\pT$<5~\GeV: 0.3--0.5~\GeV, 0.5--1~\GeV, 1--2~\GeV,
2--3~\GeV, 3--5~\GeV. Separate \pT-integrated results use 0.5<$\pT$<5~\GeV.  Per-trigger-particle yields are obtained from
the long-range ($|\deta|$>2) component of the correlation. A new
template-fitting method is applied to these yields to test for
sinusoidal modulation similar to that observed in \pPb\ collisions.

The measurements were performed using the ATLAS inner detector~(ID), minimum-bias trigger
scintillators~(MBTS), forward calorimeter (FCal), 
and the trigger and data acquisition
systems~\cite{Aad:2008zzm}. The ID detects charged particles within $|\eta|$<2.5
using a combination of silicon pixel detectors, silicon micro-strip detectors~(SCT), and 
a straw-tube transition radiation tracker~(TRT), all immersed in a 2\,T axial magnetic
field~\cite{Aad:2010bx,ATLAS_COORDINATES}. The MBTS system detects charged particles 
using two hodoscopes of counters positioned at $z = \pm 3.6~{\rm m}$.
The FCal covers 3.1<$|\eta|$<4.9 and uses tungsten and copper
absorbers with liquid argon as the active medium. Between Run 1 and
Run 2, an additional, innermost pixel layer was added to the ID and the MBTS was replaced. 

The ATLAS trigger system~\cite{Aad:2012xs} consists of a Level-1 (L1)
trigger implemented using a combination of dedicated electronics and
programmable logic, and a software-based high-level trigger (HLT).
Charged-particle tracks were reconstructed in the HLT using methods
similar to those applied in the offline analysis, allowing triggers
that select on the number of tracks with $\pT$>0.4~\GeV\ associated
with a single vertex. For the 13~\TeV\ measurements, a minimum-bias L1
trigger required one or more signals in the MBTS while the
high-multiplicity trigger (HMT) required at least 900 SCT hits and at
least 60 HLT-reconstructed tracks. For the 2.76~\TeV\ data the
minimum-bias trigger selected random crossings at L1 and applied a
threshold to the number of SCT and pixel hits in the HLT, while
several \HMT triggers were formed by applying thresholds on the total
FCal transverse energy at L1 and different thresholds on the number of
HLT-reconstructed tracks.  \HMT triggers are only used where their
multiplicity selection is more than 90\% efficient. 
The inefficiency of the \HMT triggers does not affect the measurements presented 
in this paper. This has been checked by comparing the results obtained 
with and without the HMT-triggered events, over the \Ntrk\ range where 
the \HMT is not fully efficient.

Charged-particle tracks and collision vertices are reconstructed in
the ID using algorithms that were re-optimized
between LHC Runs 1 and 2~\cite{ATL-PHYS-PUB-2015-018}.
Tracks used in the analysis are required to have $\pT$>0.3~\GeV,
$|\eta|$<2.5 and to satisfy additional selection criteria
that differ slightly between the 2.76~\cite{Aad:2012gla} and 
13~\TeV~\cite{ATLAS-CONF-2015-028} data.

\begin{figure}[!h]
\centerline{
\includegraphics[width=0.99\columnwidth]{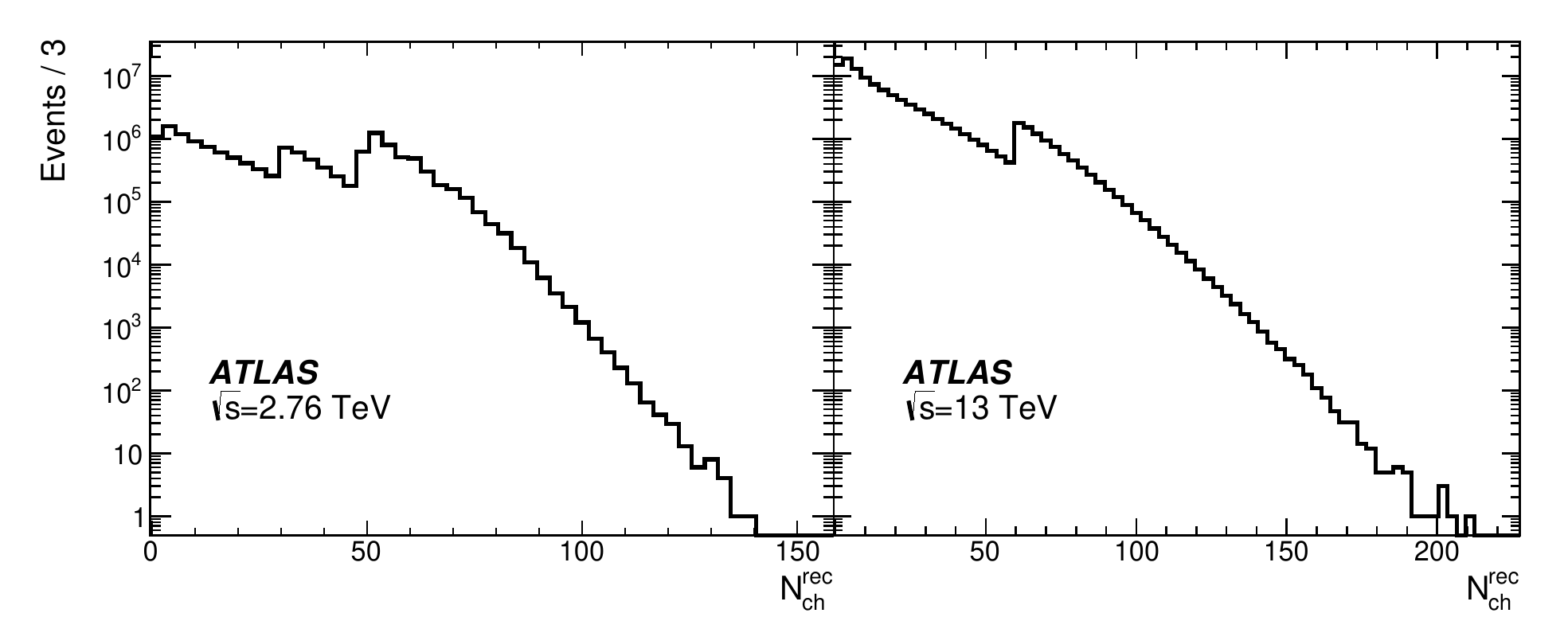}
}
\caption{
Distributions of the multiplicity, \Ntrk, of reconstructed charged particles having
$\pT$>0.4~\GeV\ for the 2.76 (left) and 13~\TeV\ (right) data used in this analysis. 
} 
\label{fig:ntrk}
\end{figure}

Events used in the analysis are required to have at least one reconstructed vertex. 
For events containing multiple vertices (pileup), only tracks associated with the 
vertex having the largest $\sum\pT^{2}$, where the sum is over all tracks 
associated with the vertex, are used. The measured charged-particle multiplicity, 
\Ntrk, is defined as the number of tracks having $\pT$>0.4~\GeV\
associated with this vertex. The distributions of \Ntrk\ are shown in Fig.\,\ref{fig:ntrk}. The structures in the 
distributions result from the different \HMT trigger thresholds. 

The efficiency, $\epsilon(\pT, \eta)$, of the track reconstruction and
track selection requirements is evaluated using simulated
non-diffractive 
\pp\ events obtained from the
\pyeight~\cite{Sjostrand:2007gs} event generator (A2
tune~\cite{mctunes}, MSTW2008LO PDFs~\cite{Sherstnev:2007nd}) that are
passed through a GEANT4~\cite{Agostinelli:2002hh}
simulation of the ATLAS detector response and
reconstructed using the algorithms applied to the data~\cite{Aad:2010ah}. The
efficiencies for the two data sets are similar, but differ due to
changes in the detector and reconstruction algorithms between Runs 1 and 2. 
In the simulated events, the efficiency reduces the measured 
multiplicity relative to the \pyeight $\pT$>0.4~\GeV\ charged-particle
multiplicity by approximately multiplicity-independent factors of
$1.18\pm0.05$ and $1.22\pm0.05$ for 13 and 2.76~\TeV\ data, respectively. 
The uncertainties in these factors result from systematic uncertainties 
in the tracking efficiencies, which are described in detail in Ref.~\cite{ATLAS-CONF-2015-028}.
Those systematic uncertainties vary with pseudorapidity between 1.1\%
(central) and 6.5\% (forward) and result from uncertainties on the material description. 

The present analysis follows methods used in previous ATLAS two-particle
correlation measurements in \mbox{Pb+Pb}  and \pPb\ collisions~\cite{Aad:2012gla,
Aad:2014lta,Aad:2012bu,Aad:2013xma,Aad:2015lwa}. Two-particle correlations 
for charged particle pairs with transverse momenta \pta\ and \ptb\  
are measured as a function of $\dphi \equiv \phiaa - \phibb$ and 
$\deta \equiv \etaaa - \etabb$, with $|\deta|$$\leq$5, determined by the 
acceptance of the ID. The particles $a$ and $b$ are conventionally referred 
to as the ``trigger'' and ``associated'' particles, respectively. The 
correlation function is defined as: 
\begin{eqnarray} 
\label{eq:ana0}
\ctwo =\frac{S(\deta,\dphi)}{B(\deta,\dphi)}\, ,
\end{eqnarray}
where $S$ and $B$ represent the same event and ``mixed event'' 
pair distributions respectively~\cite{Adare:2008ae}. When constructing $S$ and $B$,
pairs are weighted by the inverse product of their reconstruction
efficiencies $1/(\epsilon(\pta,\etaaa)\epsilon(\ptb,\etabb))$. 
Detector acceptance effects 
largely cancel in the $S/B$ ratio. 

\begin{figure}[!t]
\centerline{
\includegraphics[width=0.99\columnwidth]{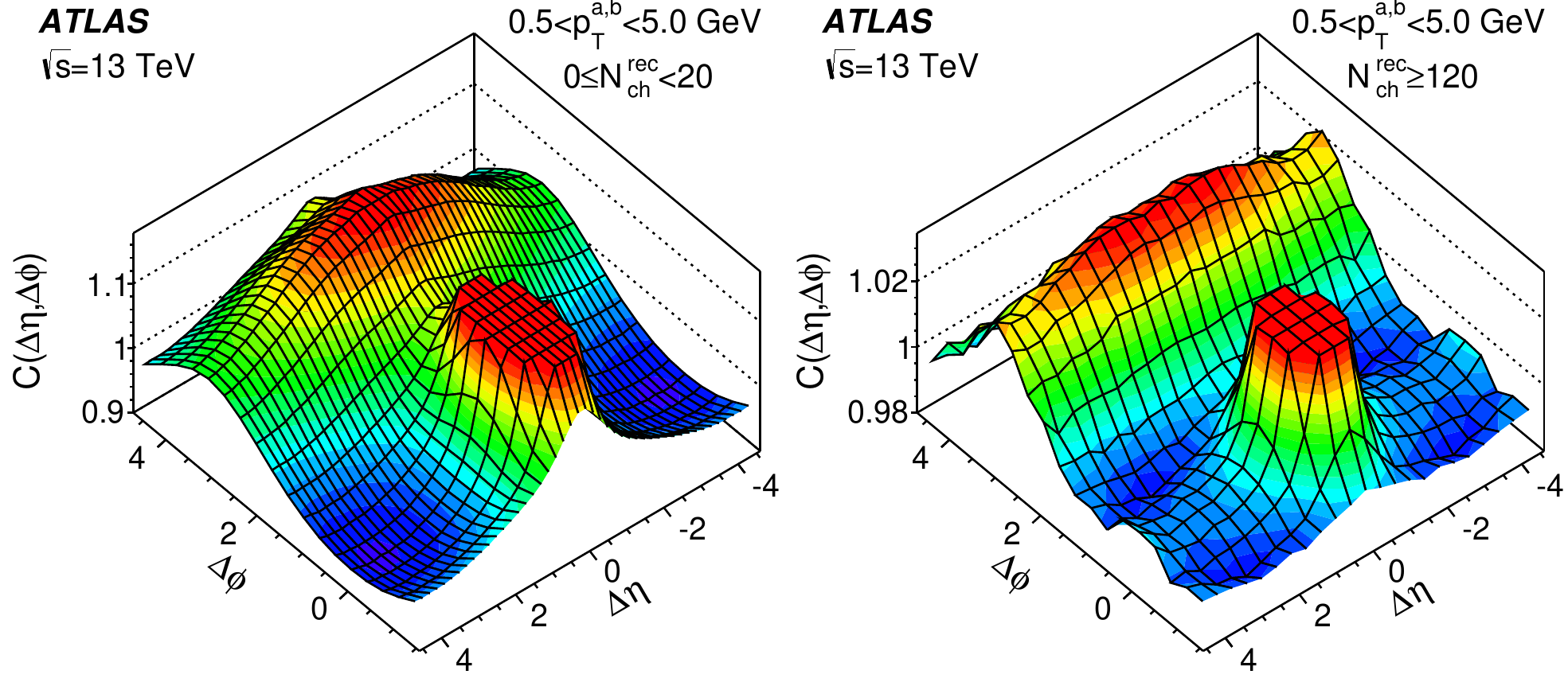}
}
\caption{
Two-particle correlation functions, \ctwo, in 13~\TeV\ \pp\ collisions
in \Ntrk\ intervals 0--20 (left) and $\geq 120$ (right) for
charged particles having 
0.5<$\ptab$<5~\GeV. The distributions have been truncated 
to suppress the peak at $\deta$=$\dphi$=0 and are shown over
$|\eta|$<4.6 to avoid statistical fluctuations at larger $|\deta|$.
}
\label{fig:c22Dmult}
\end{figure}

Examples of correlation functions in the 13~\TeV\ data are shown in Fig.\,\ref{fig:c22Dmult} for
\Ntrk intervals 0--20 (left) and $\geq$120 (right),
respectively, for 0.5<$\pTab$<5~\GeV.  
The \ctwo\ distributions have been truncated at different maximum 
values to suppress a strong peak at $\deta = \dphi = 0$ that arises
primarily from jets.
The correlation functions also show a \deta-dependent
enhancement centered at $\dphi = \pi$, which is understood to result
primarily from dijets. In the higher \Ntrk\ interval, a ridge is observed as the enhancement
near $\dphi=0$ that extends over the full \deta\ range of the measurement.

\begin{figure}[!t]
\centerline{
\includegraphics[width=0.85\columnwidth]{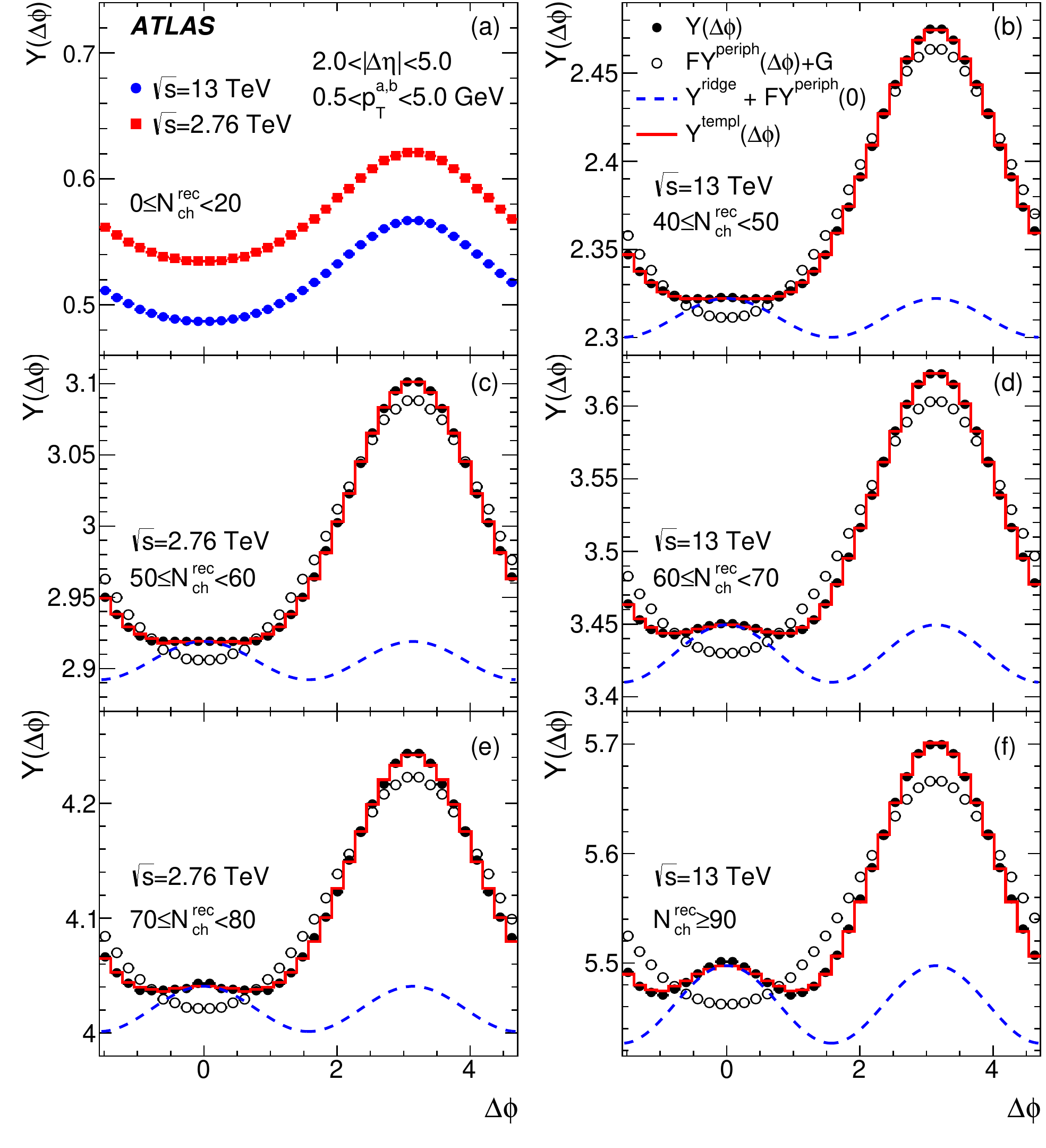}
}
\caption{
Per-trigger-particle yields, \Yphi, for 0.5<$\pTab$<5~\GeV\ 
in different \Ntrk\ intervals in 2.76 and 13~\TeV\ 
data. Panel (a): 0$\leq$$\Ntrk$<20 for both data sets. 
 Panels (c) and (e): 50--60 and 70--80 \Ntrk\ intervals for
 2.76~\TeV\ data. Panels (b), (d) and (f):  40--50, 60--70, and $\geq$90 \Ntrk\
 intervals for 13~\TeV\ data.  In panels (b)--(f),  the 
 open points and curves show different components of the template (see legend)
 that are shifted, where necessary, for presentation.
 }
\label{fig:pty}
\end{figure}

One-dimensional correlation functions, \ctwophi, are obtained by
integrating the numerator and denominator of Eq.\,\ref{eq:ana0} over
the long-range part of the correlation function, 2<$|\Delta\eta|$<5.
These are converted into ``per-trigger-particle yields,''
 \Yphi, according to \cite{Aad:2012gla,Aad:2014lta,Adare:2008ae}:  
\begin{eqnarray}
\label{eq:pty}
\Yphi  =\left( \frac{\int B(\dphi) \mathrm{d}\dphi}{ N^{a} \int \mathrm{d}\dphi}  \right) \ctwophi\, ,
\end{eqnarray}
where $N^a$ denotes the efficiency-corrected total number of trigger
particles.
Results are shown in
Fig.\,\ref{fig:pty} for selected \Ntrk\ intervals in the 13 and
2.76~\TeV\ data, for the \pTab\ ranges 0.5<$\pTab$<5~\GeV. Panel (a) 
in the figure shows \Yphi\ for 0$\leq\Ntrk$<20 for both collision
energies; these exhibit a minimum at $\dphi = 0$ and a broad peak
  at $\dphi \sim\pi$ that is understood to result primarily from
  dijets but may also include contributions from low-\pT\
  resonance decays and global momentum conservation. 
 The higher \Yphi\ values for the 2.76~\TeV\ data are due to the relative inefficiency
  of the 2.76~\TeV\ triggers for the lowest multiplicity events,
  which results in larger $\langle\Ntrk\rangle$
  for the 2.76~\TeV\ data in this \Ntrk interval. 
Panels (b), (d) and (f) show results from 13~\TeV\ data for the 40--50,
60--70, and $\geq 90$ \Ntrk\ intervals, respectively.  Panels (c) and (e)
show the results from 2.76~\TeV\ data for 50--60 and 70--80 \Ntrk\ intervals,
respectively.  With increasing \Ntrk, the minimum at $\dphi = 0$ fills
in, and a peak appears and increases in amplitude.

To separate the ridge from angular correlations present in
  low-multiplicity \pp\ collisions, a template fitting
procedure is applied to the \Yphi\ distributions.
Motivated by the peripheral subtraction method applied in
\pPb\ collisions \cite{Aad:2012gla}, the measured \Yphi\ distributions are assumed to result from a superposition of a 
``peripheral'' \Yphi distribution, scaled up by a multiplicative factor 
and a constant modulated by $\cos(2\dphi)$. The resulting template fit function,  
\begin{equation}
\YphiTempl  = F \, \YphiPer +  \YphiRidge\, ,
\label{eq:template}
\end{equation}
where 
\begin{equation}
\YphiRidge = G \left(1 + 2\vtt \cos{(2\Delta \phi)}\right)\, ,
\label{eq:template_ridge}
\end{equation}
has two free parameters, $F$ and \vtt. The coefficient, $G$, which
represents the magnitude of the combinatoric component of \YphiRidge, is fixed by requiring that
$\int_0^{\pi}{\mathrm{d}\dphi}\; Y^{\mathrm{templ}} = \int_0^{\pi}{\mathrm{d}\dphi} \;
Y$. The peripheral distribution is obtained
from the $0{\leq}\Ntrk{<}20$ interval. In the fitting
procedure, the $\chi^2$ is calculated accounting for statistical
uncertainties in both \Yphi\ and \YphiPer distributions.

Some results of the template fitting procedure are shown in panels
(b)--(f) of Fig.\,\ref{fig:pty}; a complete set of fit results is 
provided in Ref.\,\cite{self_online}. 
The scaled \YphiPer\ distributions
shifted up by $G$ are shown with open points; the
\YphiRidge\ functions shifted up by $FY^{\mathrm{periph}}(0)$ are
shown with the dashed lines; and the full fit function is shown by the
solid curves. The function in Eq.\,\ref{eq:template} successfully
describes the measured \Yphi\ distributions in all
\Ntrk\ intervals. In particular, it simultaneously describes the
ridge, which arises from an interplay of the concave \YphiPer\ and the
cosine function, the height of the peak in the \Yphi\ at $\dphi \sim
\pi$, and the narrowing of that peak which results from a negative
contribution of the $2\vtt \cos{(2\Delta \phi)}$ term in the region
near $\dphi = \pi/2$.  The agreement between
the template functions and the data allows for no significant
\Ntrk-dependent variation in the width of the dijet 
peak at $\dphi = \pi$ except for that accounted for by the sinusoidal
component of the fit function. 
Including additional $\cos{(3\dphi)}$ and $\cos{(4\dphi)}$ terms in
Eq.\,\ref{eq:template_ridge} produces changes in the extracted \vtt\ values that
are negligible compared to their statistical uncertainties. 

Previous analyses of two-particle angular correlations in \pp,
\pPb, and \mbox{Pb+Pb} collisions have traditionally relied on the ``zero
yield at minimum'' (ZYAM) hypothesis to separate the ridge from the dijet
peak at $\dphi \sim \pi$. In the ZYAM method, the ridge is
functionally defined to be $\Yphi -Y_{\mathrm{min}}$ over the
restricted  range $|\dphi| < \phimin$, where
\phimin\ is the location of the minimum of \Yphi\ and 
$Y_{\mathrm{min}} = Y(\phimin)$. However, the
\Yphi\ distributions measured in low-\Ntrk\ bins are concave in the
region near $\dphi \sim 0$.  As a result, if the ridge and dijet correlations add -- an
assumption that is implicit in all previous analyses using the ZYAM
method and is explicit in the template method used here -- then
the ZYAM method will both underestimate the ridge yield and produce
\phimin\ values that vary, unphysically, with the ridge amplitude. In
contrast, the template method used here explicitly accounts for the
concave shape of the peripheral \Yphi. Thus, the template fitting
procedure, for example, extracts a non-zero
ridge amplitude from the $\sqs = 2.76$~\TeV, $50 \leq \Ntrk \leq 60$
\Yphi\ distribution (middle left panel of Fig.\,\ref{fig:pty}) 
which is approximately flat near $\dphi \sim 0$, and would, as a
result, have approximately zero ridge signal using the ZYAM method.

Previous \pPb\ analyses used the peripheral-subtraction method, 
but applied the ZYAM procedure to the peripheral reference and, 
so, subtracted $Y(0)$ from \YphiPer.
Such a subtraction will necessarily change the \vtt\
values, and, when applied to the 13~\TeV\ data, it reduces the
measured \vtt\ by a multiplicative factor that varies from 0.4 to 0.8
over 30$\leq$$\Ntrk$<130~\cite{self_online}. 
However, if, as suggested by the data, \YphiPer\ contains not only a 
hard component, $Y^{\mathrm{hard}}(\dphi)$, but also a modulated soft component, 
\begin{equation}
\YphiPer = Y^{\mathrm{hard}}(\dphi) + G_0\left( 1 + 2 v_{2,2}^0 \cos{(2\dphi)}\right),
\end{equation}
the peripheral ZYAM method will subtract $2F G_0 v_{2,2}^0\cos{(2\dphi)}$ 
as part of the template fit, thereby reducing the
extracted \vtt. In contrast, the  procedure used in this analysis
subtracts $FG_0\left(1+2v_{2,2}^0 \cos{(2\dphi)}\right)$,  which
reduces $G$ in Eq.\,\ref{eq:template_ridge} but has less impact 
on \vtt. In particular, if $v^0_{2,2}$ 
is equal to the real \vtt\ in a given \Ntrk\ interval, there will be no bias. Since 
the measured \vtt\ is approximately \Ntrk-independent, the bias resulting from 
the presence of \vtt\ in the peripheral sample is expected to be
small. Thus, the use of the non-subtracted peripheral reference is
preferred over the more strongly biased ZYAM-subtracted reference.

\begin{figure}[!t]
\centerline{
\includegraphics[width=0.97\columnwidth]{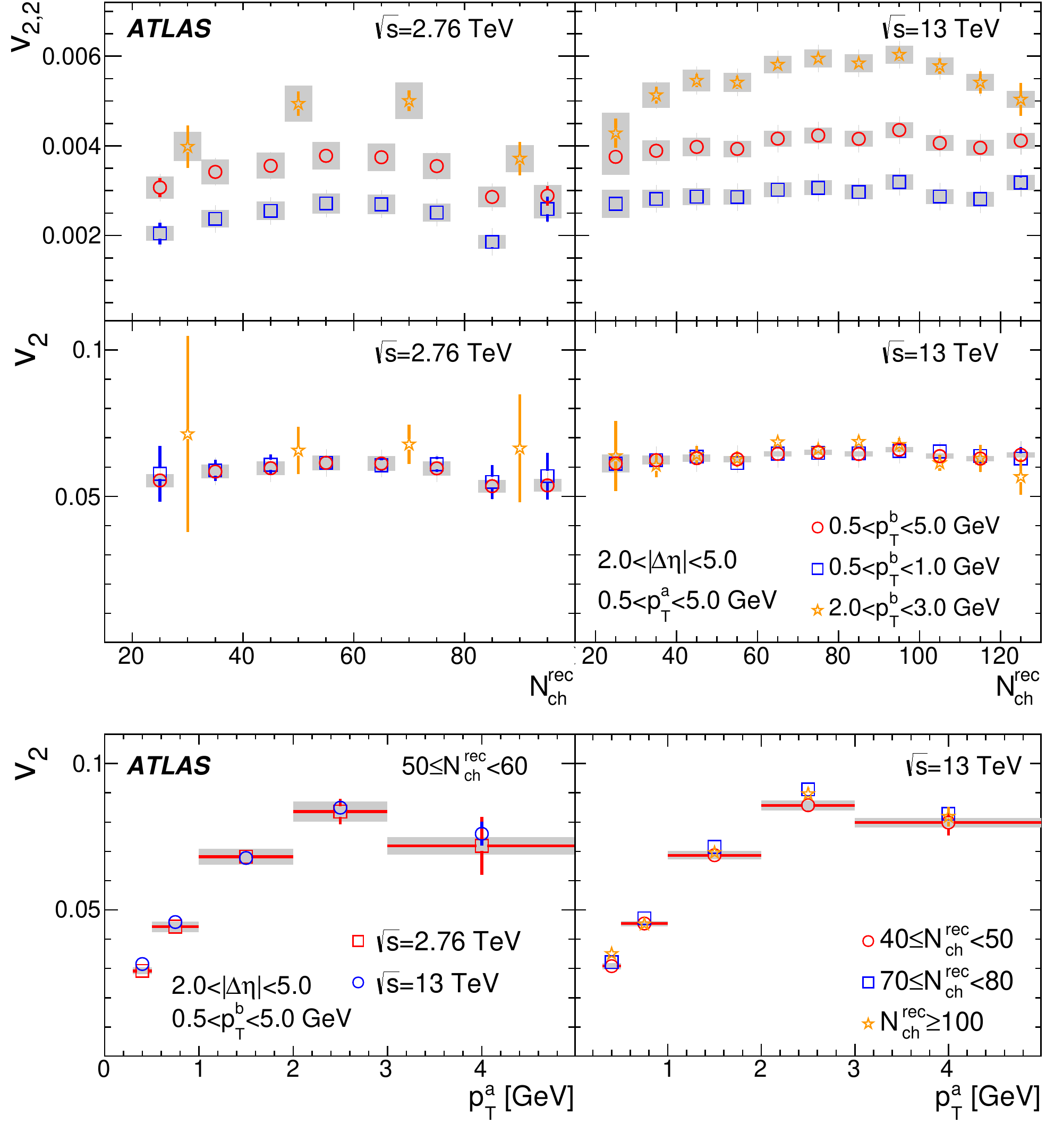}
}
\caption{
Measured \vtt\ (top) and \vtwo\ (middle) values 
versus \Ntrk\ for different \pTab\ intervals
 for 2.76 (left) and 13~\TeV\ (right) data. Results are averaged
 over \Ntrk\ bins of width 10 spanning the range 20$\leq$$\Ntrk$<100 and
 20$\leq$$\Ntrk$<130 for 2.76 and 13~\TeV\ data, respectively, except
 for the 2<$\ptb$<3~\GeV\ results for the 2.76~\TeV\ data which
 are averaged over bins of width 20.
 Measured \vtwo\ values versus \pta\ (bottom) spanning the range 0.3<$\pta$<5.0~\GeV\  for 13 and 2.76~\TeV\ data for the 50$\leq$$\Ntrk$<60
 interval (left) and for three \Ntrk\ intervals in the 13~\TeV\ data
 (right). Results are averaged over the \pta\ intervals indicated by
 horizontal error bars.
On all points, the vertical error bars indicate statistical uncertainties.
The shaded bands indicate systematic uncertainties. 
For clarity, they are only shown for the 0.5<$\ptb$<5~GeV case in the middle, 
 for 2.76~TeV data in the lower left, and for the 40$\leq$$\Ntrk$<50 case in the lower right panels.
}
\label{fig:vn}
\end{figure}

If the $\cos{(2\dphi)}$ dependence of \Yphi\
arises from modulation of the single-particle $\phi$
distributions, then \vtt\ should factorize such that $v_{2,2}
(\pta, \ptb) = v_2(\pta) v_2(\ptb)$~\cite{Aad:2012bu,Aad:2013xma,Aad:2015lwa},
where \vtwo is the $\cos(2\phi)$ Fourier coefficient of the single-particle anisotropy. 
To test this, the analysis was
performed using three \ptb\ intervals: 0.5--5, 0.5--1, and 2--3~\GeV\
with 0.5<\pta<5~\GeV; results from 2.76~\TeV\ data for the
2--3~\GeV\ interval were obtained using wider \Ntrk\ intervals to
improve statistics. Results are shown in the top panels of
Fig.\,\ref{fig:vn}; the left and right panels show 
2.76 and 13~\TeV\ data, respectively.  A significant \ptb\
dependence is seen. Separately, the same analysis was applied
requiring both \pta\ and \ptb\ to fall within the above intervals. If
factorization holds, the \vtwo\ values calculated using: 
\begin{equation}
\vtwo(\pTo) = \vtt(\pTo, \pTt) / \sqrt{\vtt(\pTt, \pTt)}\, , 
\label{eq:factortwo}
\end{equation}
where \pTo\ and \pTt\ indicate which of the three intervals, 0.5--5,
0.5--1, and 2--3~\GeV,  \pta\ and \ptb\ are required to lie within,
should be independent of \pTt. The \vtwo\ values
obtained using Eq.\,\ref{eq:factortwo} are shown in the middle panels
of Fig.\,\ref{fig:vn}. For both collision energies, the three sets
of \vtwo\ values agree within uncertainties, indicating that \vtt\ factorizes.

This analysis is sensitive to potential \Ntrk-dependent changes in the
shape of the peripheral reference. For example, the \pyeight sample shows a modest
\Ntrk-dependent change in the width of the dijet peak for small \Ntrk. Also,
the \vtt\ could vary with \Ntrk\ over the 0<$\Ntrk$<20 range. To test the 
sensitivity of the results presented here to such shape changes, the analysis 
was repeated using 0--5, 0--10, and 10--20 \Ntrk\ intervals to form \YphiPer. 
The largest resulting change in \vtt was taken as a systematic
uncertainty. The relative uncertainty varies from
6\% at $\Ntrk$=30 to 2\% for $\Ntrk$$\geq$60 in the 13~\TeV\ data, and is 
less than ${<}6\%$ for all \Ntrk\ for the 2.76~\TeV\ data. When using the
0--5 \Ntrk\ interval for \YphiPer, \vtt\ values consistent with those
shown in Fig.\,\ref{fig:vn} are measured in \Ntrk\ intervals 5--10, 10--15 and
15--20.

Potential systematic uncertainties on \vtt\ due to a residual 
\dphi\ dependence of the two-particle acceptance that does not cancel in the
$S/B$ ratio are evaluated following 
Ref.\,\cite{ATLAS:2012at} and are found to be less than 1\%.
The effect of the  uncertainty on the tracking efficiency 
on \vtt is determined to be less than 1\%. A separate systematic on \vtt\ due 
to the $\phi$ and \pT\ resolution of the charged-particle measurement is 
estimated to be 2\% (6\%) for $\pT$>0.5~\GeV\ ($\pT$<0.5~\GeV).
Events with unresolved multiple vertices decrease the
measured \vtt\ by increasing the combinatoric pedestal in \Yphi\
without increasing the modulation. The resulting systematic on \vtt\
increases with \Ntrk\, and is estimated to be less than 0.25\% and 5\% 
for the 13 and 2.76~\TeV\ data, respectively.
The combined systematic uncertainties on \vtt\ and on \vtwo\ are shown
by the shaded boxes in Fig.\,\ref{fig:vn}. The total \vtt\ systematic
uncertainty for 0.5<$\pTab$<5~\GeV\ varies between $\sim$5\% at
low \Ntrk\ to $\sim$3\% at high \Ntrk\ in the 13~\TeV\ data, while in the
2.76~\TeV\ data the uncertainty is 8\% for all \Ntrk. The systematic
uncertainty on \vtwo\ is approximately half that for \vtt. 

As shown in Fig.\,\ref{fig:vn}, the measured \vtwo\ are independent of
\Ntrk\ and are consistent between the two collision energies within
uncertainties.  The \pT\ dependence of \vtwo\ for the 50--60 \Ntrk\ interval, shown in
the bottom left panel of Fig.\,\ref{fig:vn}, is similar for both
collision energies to that previously measured in \pPb\ and \mbox{Pb+Pb}
collisions. It increases with \pT\ at low \pT, reaches a maximum between 2 and
3~\GeV, and then decreases at higher \pT. The bottom right panel of
Fig.\,\ref{fig:vn} shows the \pT\ dependence of \vtwo\ for different
\Ntrk\ intervals; no significant dependence is observed.

In summary, ATLAS has measured the multiplicity and \pT\ dependence of
two-charged-particle correlations in $\sqs$=13 and 2.76~\TeV\ \pp\
collisions at the LHC. The correlation functions at both energies
show a ridge whose strength increases with multiplicity. A
new template fitting procedure shows that the 
per-trigger-particle yields for $|\deta|$>2 are described well by a superposition
of the yields measured in a low-multiplicity interval and a constant
modulated by $\cos{(2\dphi)}$. Thus, as observed in \pPb\ collisions \cite{Aad:2012gla},
the \pp\ data presented here are compatible with both
a ``near-side'' ridge centered at $\dphi = 0$ and an ``away-side''
ridge centered at $\dphi = \pi$ that both result from a sinusoidal
component of the two-particle correlation.
The extracted Fourier
coefficients, \vtt, exhibit factorization, which is characteristic of
a global modulation of the per-event single-particle
distributions also seen in \pPb\ and \PbPb\  collisions. The amplitudes, \vtwo, 
of the single-particle modulation,
are \Ntrk-independent and agree between 2.76 and 13~\TeV\ within
uncertainties. They increase with \pT\ for $\pT\lesssim3$~\GeV\
and then decrease at higher \pT, following a trend similar to
that observed in \pPb\ and \PbPb\ collisions. These results suggest
that the ridges in \pp\ and \pPb\ collisions may arise from a similar
physical mechanism which does not have a strong \sqs\ dependence. 

\section*{Acknowledgements}
% Acknowledgements for papers with collision data
% Version 15-Sep-2015

% Standard acknowledgements start here
%----------------------------------------------
We thank CERN for the very successful operation of the LHC, as well as the
support staff from our institutions without whom ATLAS could not be
operated efficiently.

We acknowledge the support of ANPCyT, Argentina; YerPhI, Armenia; ARC, Australia; BMWFW and FWF, Austria; ANAS, Azerbaijan; SSTC, Belarus; CNPq and FAPESP, Brazil; NSERC, NRC and CFI, Canada; CERN; CONICYT, Chile; CAS, MOST and NSFC, China; COLCIENCIAS, Colombia; MSMT CR, MPO CR and VSC CR, Czech Republic; DNRF, DNSRC and Lundbeck Foundation, Denmark; IN2P3-CNRS, CEA-DSM/IRFU, France; GNSF, Georgia; BMBF, HGF, and MPG, Germany; GSRT, Greece; RGC, Hong Kong SAR, China; ISF, I-CORE and Benoziyo Center, Israel; INFN, Italy; MEXT and JSPS, Japan; CNRST, Morocco; FOM and NWO, Netherlands; RCN, Norway; MNiSW and NCN, Poland; FCT, Portugal; MNE/IFA, Romania; MES of Russia and NRC KI, Russian Federation; JINR; MESTD, Serbia; MSSR, Slovakia; ARRS and MIZ\v{S}, Slovenia; DST/NRF, South Africa; MINECO, Spain; SRC and Wallenberg Foundation, Sweden; SERI, SNSF and Cantons of Bern and Geneva, Switzerland; MOST, Taiwan; TAEK, Turkey; STFC, United Kingdom; DOE and NSF, United States of America. In addition, individual groups and members have received support from BCKDF, the Canada Council, CANARIE, CRC, Compute Canada, FQRNT, and the Ontario Innovation Trust, Canada; EPLANET, ERC, FP7, Horizon 2020 and Marie Skłodowska-Curie Actions, European Union; Investissements d'Avenir Labex and Idex, ANR, Region Auvergne and Fondation Partager le Savoir, France; DFG and AvH Foundation, Germany; Herakleitos, Thales and Aristeia programmes co-financed by EU-ESF and the Greek NSRF; BSF, GIF and Minerva, Israel; BRF, Norway; the Royal Society and Leverhulme Trust, United Kingdom.

The crucial computing support from all WLCG partners is acknowledged
gratefully, in particular from CERN and the ATLAS Tier-1 facilities at
TRIUMF (Canada), NDGF (Denmark, Norway, Sweden), CC-IN2P3 (France),
KIT/GridKA (Germany), INFN-CNAF (Italy), NL-T1 (Netherlands), PIC (Spain),
ASGC (Taiwan), RAL (UK) and BNL (USA) and in the Tier-2 facilities
worldwide.
%----------------------------------------------

\textit{Note added.}--Recently, we became aware of a related work~\cite{PhysRevLett.116.172302}.

\bibliography{ridge_PRL}
\bibliographystyle{atlasBibStyleWoTitle}

\newpage 
\input{atlas_authlist}

\end{document}

%% file: atlas_authlist.tex
% ATLAS Collaboration author list
% Data extracted on 14-Sep-2015 for paper reference HION-2015-09

\begin{flushleft}
{\Large The ATLAS Collaboration}

\bigskip

G.~Aad$^{\rm 85}$,
B.~Abbott$^{\rm 113}$,
J.~Abdallah$^{\rm 151}$,
O.~Abdinov$^{\rm 11}$,
R.~Aben$^{\rm 107}$,
M.~Abolins$^{\rm 90}$,
O.S.~AbouZeid$^{\rm 158}$,
H.~Abramowicz$^{\rm 153}$,
H.~Abreu$^{\rm 152}$,
R.~Abreu$^{\rm 116}$
% ...
\bigskip
\\
% ...
$^{11}$ Institute of Physics, Azerbaijan Academy of Sciences, Baku, Azerbaijan\\
% ...
$^{85}$ CPPM, Aix-Marseille Universit{\'e} and CNRS/IN2P3, Marseille, France\\
% ...
$^{90}$ Department of Physics and Astronomy, Michigan State University, East Lansing MI, United States of America\\
% ...
$^{107}$ Nikhef National Institute for Subatomic Physics and University of Amsterdam, Amsterdam, Netherlands\\
% ...
$^{113}$ Homer L. Dodge Department of Physics and Astronomy, University of Oklahoma, Norman OK, United States of America\\
% ...
$^{116}$ Center for High Energy Physics, University of Oregon, Eugene OR, United States of America\\
% ...
$^{151}$ Institute of Physics, Academia Sinica, Taipei, Taiwan\\
$^{152}$ Department of Physics, Technion: Israel Institute of Technology, Haifa, Israel\\
$^{153}$ Raymond and Beverly Sackler School of Physics and Astronomy, Tel Aviv University, Tel Aviv, Israel\\
% ...
$^{158}$ Department of Physics, University of Toronto, Toronto ON, Canada\\
% ...
\end{flushleft}

% Created with xml2latex.py

%% file: ridge_PRL_postRef.bbl
\begin{thebibliography}{10}
\bibitem{ATLAS_COORDINATES}
{ATLAS uses a right-handed coordinate system with its origin at the nominal
  interaction point (IP) in the center of the detector and the $z$-axis along
  the beam pipe. The $x$-axis points from the IP to the center of the LHC ring,
  and the $y$-axis points upward. Cylindrical coordinates $(r,\phi)$ are used
  in the transverse plane, $\phi$ being the azimuthal angle around the beam
  pipe. The pseudorapidity is defined in terms of the polar angle $\theta$ as
  $\eta=-\ln\tan(\theta/2)$.}
\end{thebibliography}
